\documentclass[aps,prl,preprint,groupedaddress]{revtex4}
\usepackage{graphicx}

\begin{document}

{\Large

{\bf
\ \\ \ Kondo-excitons and Auger processes in self-assembled
quantum dots}}

\vskip 0.1cm

\ \\ \ A.~O.~Govorov$^{1,2}$, K.~Karrai$^{3}$,
R.~J.~Warburton$^{4}$, and  A.~V.~Kalameitsev$^2$\ \\ \ {\it
$^1$Department of Physics and Astronomy, Ohio University, Athens,
Ohio 45701, USA \\
$^2$Institute of Semiconductor Physics, 630090 Novosibirsk, Russia \\
$^{3}$Center for NanoScience and Sektion Physik,
Ludwig-Maximilians-Universit\"{a}t, 80539 M\"{u}nchen, Germany \\
$^{4}$Department of Physics, Heriot-Watt University, Edinburgh
EH14 4AS, UK}

\vskip 0.7 cm {\bf ABSTRACT} \vskip 0.3 cm

We describe theoretically novel excitons in self-assembled quantum
dots interacting with a two-dimensional (2D) electron gas in the
wetting layer. In the presence of the Fermi sea, the optical lines
become strongly voltage-dependent. If the electron spin is
nonzero, the width of optical lines is given by $k_BT_K$, where
$T_K$ is Kondo temperature. If the spin is zero, the exciton
couples with the continuum due to Auger-like processes. This leads
to anticrossings in a magnetic field. Such states can be called
Kondo-Anderson excitons.  Some of the described phenomena are
observed in recent experiments.

\vskip 0.5 cm {\bf INTRODUCTION} \vskip 0.3 cm

Many-body phenomena and excitons in quantum dots (QDs) attract
presently much interest. The number of electrons in QDs is voltage
tunable. This makes it possible to study single electron effects.
One example of many-body phenomena is the Kondo effect induced by
the non-zero spin. So far, the Kondo effect was studied mostly in
relation to transport properties \cite{Kondo}. In optics,
Kondo-type effects were discussed with respect to nonlinear
shake-up processes in nanostructures \cite{Kondo-optics}.

Here we study theoretically novel exciton states in self-assembled
QDs coupled with delocalized electron states. If the electron spin
of the QD is non-zero, the resulting states can be called
Kondo-excitons. The Kondo temperature of these excitons is given
by the small dimensions of QDs and can be as high as $10~K$. The
Kondo effect manifests itself as peculiar, temperature-dependent
optical lines. If the exciton spin is zero, a hybridization of the
final state can occur due to Auger-like processes. Another effect
is the transition between different exciton ground states on
increasing the Fermi energy. In the transition regime, the optical
lines become strongly voltage-dependent. Some of above-mentioned
effects have been observed in optical spectra of single InAs QDs
\cite{conference-paper}.
\begin{figure}
\includegraphics[width=5.25in]{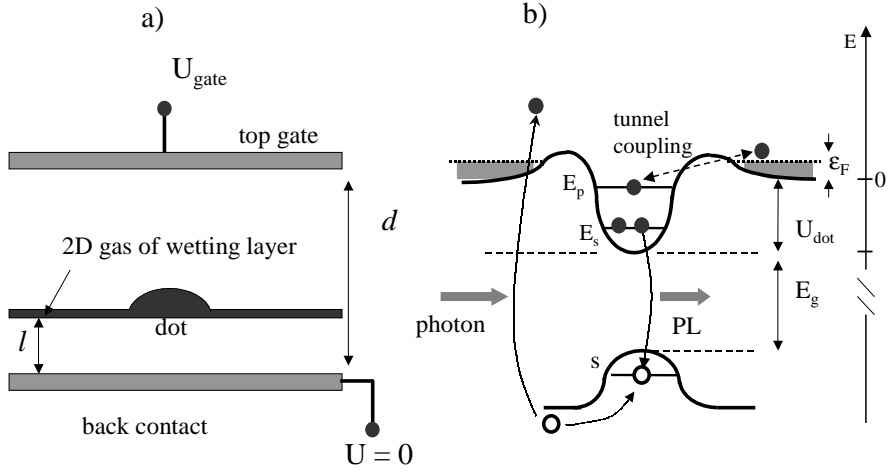}
\caption{\label{fig1} (a) Sketch of the system with quantum dots
embedded between a top gate and  back contact. By application of a
voltage, one can load single electrons in the QD and change the
Fermi energy in the wetting layer; $l\ll d$. (b) The band diagram
of a quantum dot and wetting layer.}
\end{figure}

Charged excitons $X^{n-}$, observed in experiments
\cite{conference-paper,Nature,All}, contain $n+1$ electrons and
one hole. In the exciton, the hole is optically excited; the
electrons are supplied by tunneling from the back contact and by
optical excitation (fig.1).

The voltage applied between the top and back contact, $U_{gate}$,
makes it possible to control the number of electrons in a QD. If
$U_{gate}$ is sufficiently high, electrons fill the wetting layer,
a 2D quantum well. The Fermi energy of a 2D gas in the wetting
layer is found from the conditions of equilibrium,
$\epsilon_F=(a^*_0/4d) \Delta U_{gate}$, where $a^*_0$  is the
Bohr radius and $\Delta U_{gate}$ is the voltage measured from the
point at which the wetting layer starts to load \cite{APL}. In our
model, an asymmetric QD has two bound, non-degenerate states with
energies $E_s^e$ and $E_p^e$. The indexes $s$ and $p$ originate
from electronic shells in a symmetric QD. The system is described
by the Anderson Hamiltonian,
\begin{eqnarray}
\label{AH} \hat{H}=
\hat{H}_{sp}+\hat{H}_{Coul}^{dot}+\hat{H}_{tun}, \hskip 0.2cm
\hat{H}_{tun}=\sum_{\sigma}V_k[
c_{k,\sigma}^+a_{p,\sigma}+a^+_{p,\sigma}c_{k,\sigma}], \label{H}
\end{eqnarray}
where $\hat{H}_{sp}$ is the single-particle energy;
$\hat{H}_{Coul}^{dot}$ is the operator for {\it intra-dot} Coulomb
interactions, and  $\hat{H}_{tun}$ is the operator for a tunnel
hybridization between the QD and Fermi sea; $c_{k,\sigma}$ is the
operator of a delocalized electron with a momentum $k$ and spin
$\sigma$; $\epsilon_k$ denotes the kinetic energy of delocalized
electrons; $a^+_{p,\sigma}$ is related to the $p$-state of QD.
Here we assume that the tunnel coupling occurs only between the
upper $p$-state of the QD and the Fermi sea (fig.~1). In our
simplified approach, the $p$-electron moves in a potential having
a barrier at the edges of the QD. This barrier is formed by the
short-range QD potential and Coulomb repulsion from the
$s$-electrons.

We assume that $\Delta E>U_{Coul}>E_{hyb}$, where $\Delta
E=E^e_p-E^e_s$ and $U_{Coul}$ is the intra-dot Coulomb energy, and
$E_{hyb}$ is the energy due to the hybridization between the QD
and the Fermi sea. Thus, the Coulomb interaction in a QD can be
included with perturbation theory.

The most intense lines observed in the PL spectra of QDs
\cite{Nature,All} are related to the transitions between the
$s$-states in the conduction and valence bands.  The PL spectrum
is given by a Green function:
\begin{eqnarray}
\label{Green1} I(\omega)=Re \int_0^{\infty}dt \; e^{-i\omega
t}<\hat{V}_{opt}^+(t)\hat{V}_{opt}(0)>, \hskip 0.2cm
\hat{V}_{opt}=V_{opt}(b_{s,-\frac{3}{2}}a_{s,\uparrow}+
b_{s,\frac{3}{2}}a_{s,\downarrow}+c.c.),
\end{eqnarray}
where $\omega$ is the frequency of the emitted photon and
$\hat{V}_{opt}$ is the electron-photon interaction, involving the
$s$-states of electrons and heavy holes ($J=3/2$). An averaging in
eq.~\ref{Green1} involves all initial states at a finite
temperature $T$.
\begin{figure}
\includegraphics[width=4.25in]{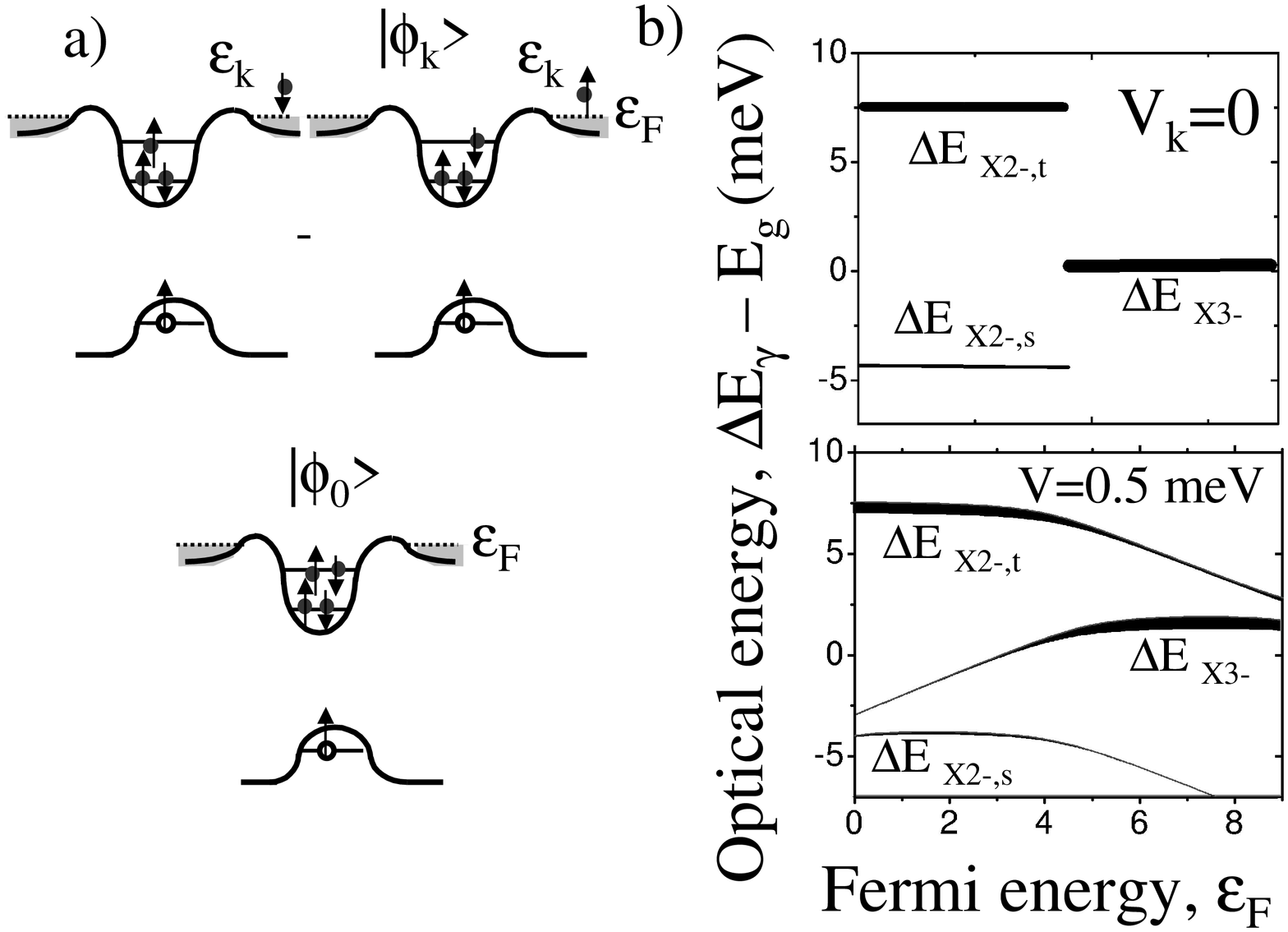}
\caption{\label{fig2} (a) Contributions to the wave functions of
excitons $X^{2-}$ and $X^{3-}$. (b) The PL spectrum as a function
of the Fermi energy with and without hybridization (upper and
lower panels). The width of the lines is proportional to the
intensity. $U_{dot}=64~meV$. }
\end{figure}
First we consider the limit $V_k\rightarrow0$ and $T\rightarrow0$.
At small Fermi energies $\epsilon_F$, the ground state of the
system is the exciton $X^{2-}$ (fig.~2a, upper part). With
increasing $\epsilon_F$, the exciton changes its charge and the
ground state becomes $X^{3-}$ (fig.~2a, lower part). The
transition Fermi energy $\epsilon_{1\rightarrow2}$ is determined
by the equation,
\begin{eqnarray}
\label{ef12} E_{X2-}+\epsilon_{1\rightarrow2}=E_{X3-},
\end{eqnarray}
where $E_{X2-}$ and $E_{X3-}$ are intra-dot exciton energies. In
fig.~2b we show the PL spectrum as a function of $\epsilon_F$ in
the absence of hybridization ($V_k=0$). At
$\epsilon_F=\epsilon_{1\rightarrow2}$, the spectrum changes
abruptly because of the ground-state transition. In the regime
$\epsilon_F<\epsilon_{1\rightarrow2}$, the spectrum has two lines,
$\Delta E_{X2-,s}$ and $\Delta E_{X2-,t}$, related to the {\it
singlet} and {\it triplet} final states \cite{Nature}. The
splitting between these lines is due to the exchange interaction
in the exciton. To calculate the optical energies, we use a model
of an anisotropic harmonic oscillator. The oscillator energies are
taken as follows: $\hbar\Omega_x^e=20~meV$,
$\hbar\Omega_y^e=25~meV$, and $\Omega_{x,y}^h=\Omega_{x,y}^e/2$,
where the upper indexes denote electrons (holes).

\vskip 0.5 cm {\bf KONDO EXCITON $X^{2-}$ WITH $S_{dot}=1/2$}
\vskip 0.3 cm

Now we focus on the interaction between the QD and 2D electrons,
assuming $V_k\neq0$ and $T=0$. In the exciton $X^{2-}$, the {\it
intra-dot electron} spin is $S_{dot}=1/2$. So, the exciton wave
function is mixed with the Fermi sea, leading to a Kondo state
with $S_e=0$, where $S_e$ is the total electron spin (fig.~2a). A
trial function for the Kondo exciton can be written as
\cite{book}:
\begin{eqnarray}
\label{WF1} \Psi_{initial}=[A_0|\phi_0>+
\sum_{k>k_F}A_k|\phi_k>]*|\uparrow,0>_h, \\
\nonumber
|\phi_0>=|\uparrow,\downarrow;\uparrow,\downarrow;\Omega>_e,
\hskip0.3cm
|\phi_k>=(\hat{c}^+_{k,\downarrow}|\uparrow,\downarrow;\uparrow,0;\Omega>_e
-\hat{c}^+_{k,\uparrow}|\uparrow,\downarrow;0,\downarrow;\Omega>_e)/\sqrt{2}
\end{eqnarray}
(fig.~2a). The first four indexes in the electron function
$|\gamma_{s\uparrow},\gamma_{s\downarrow};\gamma_{p\uparrow},
\gamma_{p\downarrow};\Omega>_e$ represent the occupations of the
QD;  $\Omega$ denotes all states of the Fermi sea with $k<k_F$,
where $k_F$ is the Fermi momentum. The function $|\uparrow,0>_h$
describes the hole with $J=+3/2$. Note that the electronic part of
the function (\ref{WF1}) is a singlet state with $S_e=0$.  By
using the Hamiltonian (\ref{AH}), we find the unknown coefficients
in eq.~\ref{WF1} and the ground state energy $E_{i}=
E_{X2-}+\epsilon_F-\delta$. The lowering of energy $\delta$ plays
the role of a Kondo temperature, $T_K$. If
$E_p^e+U_p-\epsilon_F>\delta$, we obtain
\begin{eqnarray}
\label{TK}
k_BT_{K}=\delta=(D-\epsilon_F)e^{-\frac{\pi(E_p^e+U_p-\epsilon_F)}{2\Delta}},
\end{eqnarray}
\cite{book}. Here $\Delta=\pi V_0^2\rho_0$ is the broadening and
$\rho_0$ is the 2D density of states (DOS). $U_p$ is the
Coulomb-blockade energy for the $p$-state; $D$ is the cut-off
parameter for the tunnel matrix element $V_k$: $V_k=V_0$ for
$\epsilon<D$ and $0$ elsewhere \cite{D}. For example, $T_{K}\sim
7-14~K$ for $\epsilon_F=2-3.6~meV$, $\Delta=1~meV$, and
$D\sim30~meV$.

The PL spectrum can be calculated by using eqs.~\ref{Green1} and
\ref{WF1}. In the regime $\epsilon_F<\epsilon_{1\rightarrow2}$,
the PL spectrum contains three lines. The main lines are singlet
and triplet states of $X^{2-}$. The third structure is a shake-up
line due to $X^{3-}$. The most intense, triplet line has an
asymmetric shape \cite{govorov}:
\begin{eqnarray}
\label{spectr1} I_{X2-,t}=\frac{3}{4}V_{opt}^2 \frac{2\Delta
A_0^2}{\pi}
\int_0^{D-\epsilon_F}d\epsilon\frac{1}{(\epsilon+k_BT_{K})^2}
Re\frac{-i}{\hbar\omega-\Delta E'_{X2-,t}+\epsilon-i\gamma_t},
\end{eqnarray}
where $\Delta E'_{X2-,t}=\Delta E_{X2-,t}-k_BT_{K}$ is the
renormalized exciton energy, $\Delta E_{X2-,t}$ is the optical
energy in the absence of the 2D gas, and $\gamma_{t}$ describes
the phonon-induced broadening of the final state. Typically, the
broadening of the triplet intra-dot state $\gamma_t$ is small,
since the energy relaxation in this case involves a spin-flip
process.
\begin{figure}
\includegraphics[width=4.25in]{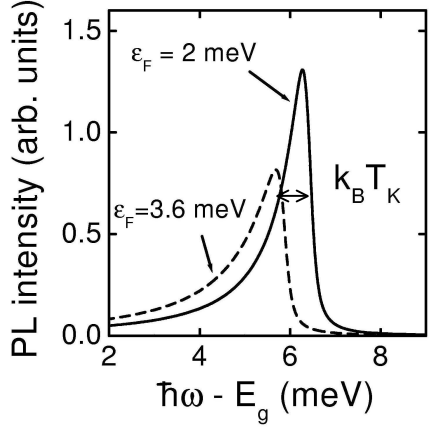}
\caption{\label{fig3} PL line of $X^{2-,t}$ exciton;
$U_{dot}=64~meV$ and $\Delta=1~meV$.  }
\end{figure}
If $KT_{K}\gg\gamma_t$, the triplet line is asymmetric and its
width is equal to $KT_{K}$ (fig.~3). The shape and width of the
line originate from the peak of the spectral density of states
(SDOS) at the Fermi level in the initial Kondo state given by
eq.~\ref{WF1}. The line shape and peak position in fig.~4 depend
on the Fermi energy and, therefore, are voltage-dependent.
Besides, we think that the line width is strongly
temperature-dependent, because the Kondo peak in the SDOS vanishes
as the temperature increases \cite{book}.

\vskip 0.4 cm {\bf EXCITON $X^{3-}$ WITH $S_{dot}=0$: AUGER
PROCESSES} \vskip 0.3 cm

In the case $\epsilon_F>\epsilon_{1\rightarrow2}$, the main peak
in the spectrum comes from the exciton $X^{3-}$ where there are
two completely filled electron levels. Its intra-dot electron spin
is zero. Thus, the Kondo effect does not occur in the initial
state. However, after photon emission, the final state of the
exciton $X^{3-}$ is excited and can interact with the continuum of
states due to the Auger-like process induced by the intra-dot
Coulomb interaction (fig.~4a). In such a process, the final state
is a linear combination of the wave functions
$|f_0>=|\uparrow,0;\uparrow,\downarrow;\Omega>$ and
$|f_{1,k}>=\hat{c}_{k}^+|\uparrow,\downarrow;0,0;\Omega>$, where
$k>k_F$ \cite{eh}. The energies of the above states can be written
as $\epsilon_0=E_0^{dot}$ and
$\epsilon_{1,k}=E_1^{dot}+\epsilon_k$, where $E_0^{dot}$ and
$E_1^{dot}$ are intra-dot energies and $\epsilon_k$ is the kinetic
energy. The parameter $\delta\epsilon_a=E_0^{dot}-E_1^{dot}$
determines the excess kinetic energy, when one of the QD electrons
is excited into the continuum (fig.~4a).

By using non-crossing diagrams (fig.~4b), we write the Green
function (\ref{Green1}) in the form
\begin{eqnarray}
\label{Green2}
I(\omega)=V_{opt}^2Re[\frac{-i}{\tilde{\omega}-i0-\Sigma_1}],
\hskip 0.4 cm
\Sigma_1=\int_0^{\infty}\frac{W_{\epsilon}^2\rho(\epsilon)[1-f_\epsilon]
d\epsilon}{\tilde{\omega}-\delta\epsilon_a+
\epsilon-i0-\int_0^{\infty}\frac{\rho(\epsilon')W_{\epsilon'}^2
f_{\epsilon'}d\epsilon' }{\tilde{\omega}+\epsilon-\epsilon'-i0} },
\end{eqnarray}
where $\Delta E_{X3-}$ is the optical-line energy,
$\tilde{\omega}=\omega-\Delta E_{X3-}$, $\rho$ is the DOS and
$f_\epsilon$ is the Fermi-distribution function; the Auger matrix
element $W_{\epsilon_k}=<f_0|U_{Coul}|f_{1,k}>$, where $U_{Coul}$
is the Coulomb interaction. For the parameter $W_\epsilon$, we
assume: $W_\epsilon=W_0$ if $\epsilon<D'$ and $0$ elsewhere. In
our approach, the final states in the PL process are described by
the non-interacting Anderson model, in which the Auger matrix
element plays the role of tunnel element $V_k$.
\begin{figure}
\includegraphics[width=4.25in]{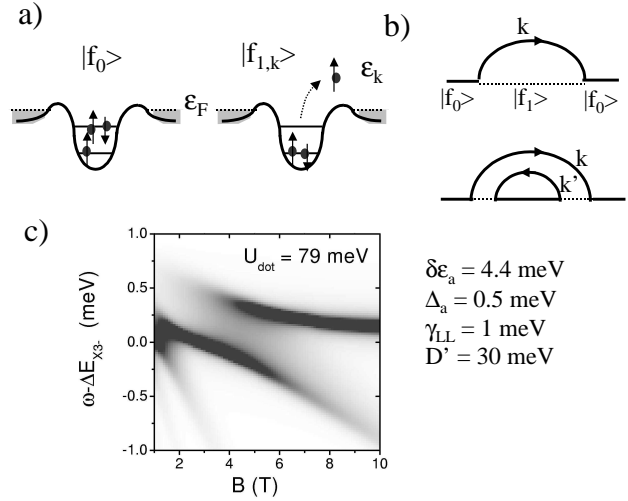}
\caption{\label{fig4} (a) Final intra-dot states of the exciton
$X^{3-}$. (b) Non-crossing diagrams. (c) The PL spectrum of
$X^{3-}$ in the magnetic field; $\gamma_{LL}$ is the Landau-level
broadening. }
\end{figure}

{\bf (i) Interaction with the continuum: empty wetting layer}\\ We
now assume that the QD is sufficiently deep and the wetting layer
is empty; $U_{dot}=79~meV$. In this case, we can put
$f(\epsilon)=0$.
In the absence of a magnetic field, the PL peak (eq.~\ref{Green2})
is close to a Lorentzian. The broadening is given by $\Delta_a=\pi
W_0^2\rho_0$. The estimated Auger-broadening $\Delta_a$ can be as
high as 2 $meV$ \cite{govorov,Bastard}. Note that this mechanism
of broadening is found in experimental spectra of $X^{3-}$
\cite{conference-paper}. In the magnetic field $B$, the DOS
becomes strongly modulated, due to formation of Landau levels,

\begin{eqnarray}
\label{LL}
\rho(\epsilon)=\rho_0\frac{\hbar\omega_c}{\sqrt{\pi}\gamma_{LL}}
\sum_{n}e^{-\frac{(\epsilon-\epsilon_n)^2}{\gamma_{LL}^2}},
\end{eqnarray}
where $\gamma_{LL}$ is the Landau level broadening, 
$\epsilon_n=(n+1/2)\hbar\omega_c$ is the Landau level energy, and
$\omega_c$ is the cyclotron frequency. In this case, the optical
energies demonstrate anticrossings between the terms:
$\tilde{\omega}=0$ and
$\tilde{\omega}-\delta\epsilon_a+(n+1/2)\hbar\omega_c=0$
(fig.~4c). The anticrossing strength increases with the magnetic
field as $\sqrt{\Delta_a\hbar\omega_c}$. This is due an increase
of the density of states. \\

{\bf (ii) Spectrum of $X^{3-}$ in the presence of a 2d gas}\\ If
$T=0$, $B=0$ and $\delta\epsilon_a-\epsilon_F>\Delta_a$, the line
is again close to a Lorentzian with a high-energy cut-off at the
frequency $\tilde{\omega}=\delta\epsilon_a-\epsilon_F$. If
$B\neq0$, the PL spectrum in the presence of 2D gas is more
complex due to electron-hole excitations in the Fermi sea.

\vskip 0.4 cm {\bf TRANSITION FROM $X^{2-}$ TO $X^{3-}$} \vskip
0.3 cm

When $\epsilon_F\sim\epsilon_{1\rightarrow2}$, the ground state
$X^{2-}$ changes to the state $X^{3-}$. In the transition regime,
the exciton strongly couples with the Fermi sea. Hence, we can
expect that the optical energies should strongly depend on
$\epsilon_F$. In fig.~2b, we show this effect by using a
simplified, zero bandwidth model \cite{govorov,model}.   We can
see that the energies $\Delta E_{X2-,t(s)}$ decrease with
$\epsilon_F$ and, thus, are voltage-dependent. Such a behavior has
been observed in recent experiments \cite{conference-paper}.

\vskip 0.4 cm
{\bf CONCLUSIONS} \vskip 0.3 cm

Novel exciton states can appear in QDs interacting with the Fermi
gas. The specific feature of these states is a ``cloud" of
electrons around the charged exciton. The radius of the
Kondo-Anderson excitons is given by the size of the electron cloud
and can be greater than the size of QD.  We describe several
striking manifestations of such excitons in PL spectra. Some of
described phenomena are seen in experimental spectra of InAs QDs
\cite{conference-paper}.


\begin{thebibliography}{10}
\bibitem{D}
The matrix element $V_k$ decreases with the kinetic energy
$\epsilon_k$. The cut-off parameter $D$ can be estimated as
$\hbar^2/m^*a^2$, where $a$ and $m^*$ are the QD lateral size and
electron mass, respectively.
\bibitem{model}
In the zero bandwidth model, we replace the Fermi sea by a single
``delocalized" state of energy $\epsilon_F$ \cite{book}.
\end{thebibliography}
\end{document}